\documentclass[12pt]{article}
\usepackage{amsfonts}
\usepackage{cite}
\usepackage{amsmath}
\usepackage{amssymb}
\usepackage{stackrel}
\usepackage{color}

\providecommand{\U}[1]{\protect\rule{.1in}{.1in}}
\makeatletter
\@addtoreset{equation}{section}
\makeatother

\input{tcilatex}
\begin{document}

\begin{titlepage}
\begin{center}
\renewcommand{\thefootnote}{\fnsymbol{footnote}}
{\Large{\bf Special Geometry and Space-time Signature}}
\vskip1cm
\vskip 1.3cm
W. A. Sabra
\vskip 1cm
{\small{\it
Centre for Advanced Mathematical Sciences and Physics Department\\
American University of Beirut\\ Lebanon \\
ws00@aub.edu.lb}}

\vskip 1.3cm

\end{center}
\bigskip
\begin{center}
{\bf Abstract}
\end{center}
 We construct ${\cal N}=2$  four and five-dimensional supergravity theories coupled to vector multiplets in 
various space-time signatures $(t,s)$, where $t$ and $s$ refer, respectively, to the number of time and 
spatial dimensions. The five-dimensional supergravity theories, $ t+s=5$, are constructed  by investigating  the
 integrability conditions arising from Killing spinor equations. 
The five-dimensional supergravity theories can also be obtained by reducing Hull's eleven-dimensional 
supergravities on a Calabi-Yau threefold. 
The dimensional reductions of the five-dimensional supergravities on space and time-like circles 
produce ${\cal N}=2$ four-dimensional supergravity theories with signatures $(t-1,s)$ and $(t,s-1)$ exhibiting 
projective special (para)-Kähler geometry.
\end{titlepage}\ 

\section{Introduction}

The study of the geometry of the scalar manifolds of Euclidean $\mathcal{N}%
=2 $ vector and hypermultiplets with or without coupling to supergravity has
recently been considered in \cite{mohaupt1, mohaupt2, mohaupt3, mohaupt4}.
Our present work will only focus on theories with vector multiplets coupled
to supergravity. In the standard supergravity theories with Lorentzian
signature, it is well known that the scalar manifold is described by a
projective special K\"{a}hler manifold in four dimensions and by a
projective special real manifold in five dimensions. In \cite{mohaupt3},
four-dimensional Euclidean supergravity theories were obtained by
dimensionally reducing the five-dimensional Lorentzian theories of \cite{GST}
over a time-like circle. It was established that the scalar geometries of
the four-dimensional Euclidean vector multiplets can be obtained by
replacing complex structures by para-complex structures. Four-dimensional
Euclidean supergravity theories were also obtained as a dimensional
reduction of the Euclidean ten-dimensional supergravity on a Calabi-Yau
three-fold \cite{tenred}. The Killing spinor equations of the Euclidean
four-dimensional supergravities were also obtained in \cite{ks} and their
gravitational solutions admitting Killing spinors were analysed in \cite%
{ginst}.

The theories of $\mathcal{N}=2,$ five-dimensional Euclidean vector
multiplets coupled to supergravity has recently been constructed in \cite{e5}%
. The Lagrangian of the Euclidean theory is the same as in the Lorentzian
theory except that the gauge fields terms appear with the opposite sign.
Multi-centered solutions of the gauged versions of these theories were
recently studied in \cite{multi}. The dimensional reduction of the
five-dimensional Euclidean theory on a circle produces the $\mathcal{N}=2$
four-dimensional Euclidean supergravity of \cite{mohaupt3} but with the
non-conventional signs of the gauge terms.

In this work, our aim is to obtain $\mathcal{N}=2$ four and five-dimensional
supergravity theories coupled to vector multiplets in various space-time
signatures $\left( t,s\right) ,$ where $t$ and $s$ refer, respectively, to
the number of time and spatial dimensions. Space-times with various
signatures are of mathematical and physical interest. For instance, spaces
with $(2,2)$ signature have applications to string theory, M-theory,
cosmology and twistor theory \cite{bars}. Moreover, the $(2,2)$ theory and
its solutions without matter fields, have been considered in \cite{brdun}.
More recently, a classification of solutions with Killing spinors for the $%
(2,2)$ Einstein-Maxwell theory with a cosmological constant was given in 
\cite{kn}. We organise our work as follows. In Sec. 2, the five-dimensional
supergravity theories are constructed through the analysis of the
integrability conditions arising from generalised Killing spinor equations.
The theories with various signatures are then obtained by reducing Hull's
eleven-dimensional supergravity \cite{Hull} on Calabi-Yau threefold. In Sec.
3, we obtain new $\mathcal{N}=2$ four-dimensional supergravity theories via
the dimensional reductions of the five-dimensional supergravities on space
and time-like circles. In particular, we obtain $\mathcal{N}=2$ supergravity
with signature $(2,2)$ with scalar manifold described by a projective
special para-K\"{a}hler manifold. We also obtain the Killing spinor
equations for the reduced four-dimensional $\mathcal{N}=2$ supergravity
theories. We end with a summary of our results.

\section{$(t,s)$ Five-Dimensional Supergravity}

We start our analysis with the original theory of $\mathcal{N}=2$, $D=5$
supergravity theory coupled to Abelian vector multiplets constructed in \cite%
{GST}. The theory contains the gravity multiplet and $n$ vector multiplets
and its bosonic Lagrangian is given by

\begin{equation}
\hat{\mathbf{e}}^{-1}\hat{\mathcal{L}_{5}}=\frac{1}{2}\hat{R}-\frac{1}{2}%
G_{ij}\partial _{\hat{m}}h^{i}\partial ^{\hat{m}}h^{j}-\frac{1}{4}%
G_{ij}F^{i}{}_{\hat{m}\hat{n}}F^{j}{}^{\hat{m}\hat{n}}+\frac{\hat{\mathbf{e}}%
^{-1}}{48}\,C_{ijk}\epsilon ^{\hat{n}_{1}\hat{n}_{2}\hat{n}_{3}\hat{n}_{4}%
\hat{n}_{5}}F^{i}{}_{\hat{n}_{1}\hat{n}_{2}}F^{j}{}_{\hat{n}_{3}\hat{n}%
_{4}}A^{k}{}_{\hat{n}_{5}},  \label{act}
\end{equation}%
where $C_{ijk}$ are real constants symmetric in $i,j,k$. The dynamics of (%
\ref{act}) is encoded in the cubic potential 
\begin{equation}
\mathcal{V}={\frac{1}{6}}C_{ijk}h^{i}h^{j}h^{k}\,,
\end{equation}%
where the very special coordinates $h^{i}$ are functions of the $n$ real
scalar fields belonging to the vector multiplets. The scalar manifold is
described by the very special geometry 
\begin{equation}
\mathcal{V}={1.}
\end{equation}%
The gauge coupling metric takes the form%
\begin{equation}
G_{ij}=-{\frac{1}{2}}\left( {\partial _{h^{i}}}{\partial _{h^{j}}}(\ln 
\mathcal{V})\right) _{\mathcal{V}=1}=\frac{1}{2}\left( {9}%
h_{i}h_{j}-C_{ijk}h^{k}\right) ,  \label{gc}
\end{equation}%
where the dual fields $h_{i}$ are given by 
\begin{equation}
h_{i}={\frac{1}{6}}C_{ijk}h^{j}h^{k}.  \label{d}
\end{equation}%
The Killing spinor equations arising from the vanishing of the fermionic
fields and their supersymmetry transformations can be written in the form 
\footnote{%
Our conventions are as follows: The Clifford algebra is $\{{\Gamma
^{a},\Gamma ^{b}}\}=2\eta ^{ab}$. The covariant derivative on spinors is ${{{%
\hat{D}}}}_{\hat{m}}=\partial _{\hat{m}}+{\frac{1}{4}}\omega _{\hat{m},ab}{%
\Gamma }^{ab}$ where $\omega _{\hat{m},ab}$ is the spin connection. Finally,
antisymmetrization is with weight one, so ${\Gamma }^{{a}_{1}{a}_{2}\cdots {%
a_{n}}}={\frac{1}{n!}\Gamma }^{[{a_{1}}}{\Gamma }^{{a_{2}}}\cdots {\Gamma }^{%
{a_{n}}]}$.}

\begin{align}
\left[ {{{\hat{D}}}}_{\hat{m}}+{\frac{i}{8}}h_{i}\left( \Gamma _{\hat{m}}{}^{%
{\hat{n}}_{1}{\hat{n}}_{2}}-4\delta _{{\hat{m}}}^{{\hat{n}}_{1}}\Gamma ^{{%
\hat{n}}_{2}}\right) F_{{\hat{n}}_{1}{\hat{n}}_{2}}^{i}\right] {\hat{%
\varepsilon}}& =0,  \notag \\
\left[ i\left( F^{i}-h^{i}h_{j}F^{j}\right) _{{\hat{n}}_{1}{\hat{n}}%
_{2}}\Gamma ^{{\hat{n}}_{1}{\hat{n}}_{2}}-2{\partial }_{\hat{m}}h^{i}\Gamma
^{\hat{m}}\right] {\hat{\varepsilon}}& =0.  \label{ks51}
\end{align}

In what follows we obtain five-dimensional theories with various space-time
signatures. One method to do so is through the analysis of the integrability
of the Killing spinor equations. We start by allowing for a slight
modification of the Killing spinor equations and write

\begin{align}
\left[ {{{\hat{D}}}}_{\hat{m}}+{\frac{\alpha }{8}}h_{i}\left( \Gamma _{\hat{m%
}}{}^{{\hat{n}}_{1}{\hat{n}}_{2}}-4\delta _{{\hat{m}}}^{{\hat{n}}_{1}}\Gamma
^{{\hat{n}}_{2}}\right) F_{{\hat{n}}_{1}{\hat{n}}_{2}}^{i}\right] {\hat{%
\varepsilon}}& =0,  \notag \\
\left[ \alpha \left( F^{i}-h^{i}h_{j}F^{j}\right) _{{\hat{n}}_{1}{\hat{n}}%
_{2}}\Gamma ^{{\hat{n}}_{1}{\hat{n}}_{2}}-2{\partial }_{\hat{m}}h^{i}\Gamma
^{\hat{m}}\right] {\hat{\varepsilon}}& =0.
\end{align}%
After some calculation one can derive the following integrability condition%
\begin{eqnarray}
&&2\alpha \left[ \left( {{{\hat{D}}}}^{\hat{\nu}}\left( G_{ij}F_{\hat{\nu}%
\hat{\lambda}}^{j}\right) -h_{i}h^{j}{{{\hat{D}}}}^{\hat{\nu}}\left(
G_{jl}F_{\hat{\nu}\hat{\lambda}}^{l}\right) \right) \Gamma ^{\hat{\lambda}}+%
\frac{\alpha }{16}\left( h_{i}C_{jkl}h^{l}-C_{ijk}\right) F_{\hat{\lambda}%
_{1}\hat{\lambda}_{2}}^{j}F_{\hat{\lambda}_{3}\hat{\lambda}_{4}}^{k}\Gamma ^{%
\hat{\lambda}_{1}\hat{\lambda}_{2}\hat{\lambda}_{3}\hat{\lambda}_{4}}\right] 
{\hat{\varepsilon}}  \notag \\
&&+\alpha ^{2}F_{\hat{\lambda}_{1}\hat{\lambda}_{2}}^{j}F^{k\hat{\lambda}_{1}%
\hat{\lambda}_{2}}\left( 9h_{i}h_{j}h_{k}-\frac{1}{4}C_{jkl}h^{l}h_{i}+\frac{%
1}{4}C_{ijk}-\frac{3}{2}C_{ijm}h^{m}h_{k}\right) {\hat{\varepsilon}}  \notag
\\
&&+\left( 3{{{\hat{D}}}}^{\hat{\nu}}{{{\hat{D}}}}_{\hat{\nu}}h_{i}+\frac{9}{2%
}h_{i}{{{\hat{D}}}}_{\hat{\nu}}h_{j}{{{\hat{D}}}}^{\hat{\nu}}h^{j}-\frac{1}{2%
}C_{ijk}{{{\hat{D}}}}_{\hat{\nu}}h^{j}{{{\hat{D}}}}^{\hat{\nu}}h^{k}\right) {%
\hat{\varepsilon}}  \notag \\
&=&0.  \label{integral}
\end{eqnarray}%
The vanishing of the second and third lines in the above equation constitute
the equations of motion for the scalar fields in a theory where the gauge
kinetic terms coefficient is $\frac{\alpha ^{2}}{4}$. Assuming that this is
the case, then (\ref{integral}) reduces to

\begin{equation}
\left[ {{{\hat{D}}}}^{\hat{\nu}}\left( G_{ij}F_{\hat{\nu}\hat{\lambda}%
}^{j}\right) \Gamma ^{\hat{\lambda}}-\frac{\alpha }{16}C_{ijk}F_{\hat{\lambda%
}_{1}\hat{\lambda}_{2}}^{j}F_{\hat{\lambda}_{3}\hat{\lambda}_{4}}^{k}\Gamma
^{\hat{\lambda}_{1}\hat{\lambda}_{2}\hat{\lambda}_{3}\hat{\lambda}_{4}}%
\right] {\hat{\varepsilon}}=0.  \label{ing}
\end{equation}%
If we modify the action (\ref{act}) to take the form 
\begin{equation}
\hat{\mathbf{e}}^{-1}\hat{\mathcal{L}_{5}}=\frac{1}{2}\hat{R}-\frac{1}{2}%
G_{ij}\partial _{\hat{m}}h^{i}\partial ^{\hat{m}}h^{j}+\frac{\alpha ^{2}}{4}%
\left( G_{ij}F^{i}{}_{\hat{m}\hat{n}}F^{j}{}^{\hat{m}\hat{n}}-\frac{\hat{%
\mathbf{e}}^{-1}}{12}\,C_{ijk}\epsilon ^{\hat{n}_{1}\hat{n}_{2}\hat{n}_{3}%
\hat{n}_{4}\hat{n}_{5}}F^{i}{}_{\hat{n}_{1}\hat{n}_{2}}F^{j}{}_{\hat{n}_{3}%
\hat{n}_{4}}A^{k}{}_{\hat{n}_{5}}\right) ,\   \label{act23}
\end{equation}%
then the equations of motion for the gauge fields derived from (\ref{act23}%
)\ are given by 
\begin{equation}
{{{\hat{D}}}}^{\hat{\nu}}\left( G_{ij}F_{\hat{\nu}\hat{\lambda}}^{j}\right) +%
\frac{1}{16}C_{ijk}\epsilon ^{\hat{\lambda}_{1}\hat{\lambda}_{2}\hat{\lambda}%
_{3}\hat{\lambda}_{4}}F_{\hat{\lambda}_{1}\hat{\lambda}_{2}}^{j}F_{\hat{%
\lambda}_{3}\hat{\lambda}_{4}}^{k}=0.  \label{gem}
\end{equation}

In theories with $(1,4)$, $(3,2)$ and $(5,0)$ signatures (odd numbers of
time dimensions), (\ref{ing}) is consistent with (\ref{gem}) for the
standard sign of the gauge terms, i.e., $\alpha =i.$ For the mirror theories
with signatures $(4,1),(2,3)$ and $(0,5)$, consistency implies $\alpha =-1$
and thus the opposite sign of the gauge terms.

Five-dimensional $\mathcal{N}=2$ supergravity theories with $\left(
1,4\right) $ signature can be obtained via the dimensional reduction of
eleven-dimensional supergravity with the bosonic action \cite{cjs}

\begin{equation}
S_{11}=\int_{M_{11}}\frac{1}{2}R\tilde{\ast}1-\frac{1}{4}F_{4}\wedge \tilde{%
\ast}F_{4}-\frac{1}{12}C_{3}\wedge F_{4}\wedge \tilde{\ast}F_{4}  \label{11d}
\end{equation}%
and signature $\left( 1,10\right) $ on a Calabi-Yau three-fold, $CY_{3}$ 
\cite{cad}. Here $F_{4}=dC_{3}$ and $C_{3}$ is a $3$-form. The
eleven-dimensional space-time manifold decomposes into $M_{11}=CY_{3}\times
M_{5}$, where $M_{5}$ is a Lorentzian five-dimensional manifold. Some useful
details on the mathematics of $CY_{3}$ as well as the reduction on $CY_{3}$
can be found for example in \cite{Bod, Fer, Candelas, gr}.

We shall briefly present the basics of the reduction relevant to our
discussion. One considers the deformations of the $CY_{3}$ metric that
preserve the $SU(3)$ holonomy. These are the zero modes of the internal wave
operator which correspond to deformations of the K\"{a}hler class and the
complex structure. Ignoring the complex structure moduli, fluctuations of
the $CY_{3}$ metric are then expanded as

\begin{equation}
\delta g_{A\bar{B}}=-iV_{A\bar{B}}^{i}\delta q^{i}
\end{equation}%
where $q^{i}$ are the K\"{a}hler moduli taken to depend on the coordinates
of $M_{5}$ and

\begin{equation}
V^{i}=V_{A\bar{B}}^{i}d\xi ^{A}\wedge d\bar{\xi}^{\bar{B}},\text{ \ \ \ \ \ }%
i=1,\ldots ,h_{1,1},\text{ }
\end{equation}%
are the basis of $h_{(1,1)}$ harmonic forms, $\xi ^{A}$ represent the three
complex coordinates of $CY_{3}.$ Note that the K\"{a}hler form and the
volume of $CY_{3}$ are given by 
\begin{eqnarray}
J &=&ig_{A\bar{B}}d\xi ^{A}\wedge d\bar{\xi}^{\bar{B}}=q^{i}V^{i},  \notag \\
\mathcal{\tilde{V}} &=&\frac{1}{3!}\int_{CY_{3}}J\wedge J\wedge J=\frac{1}{6}%
C_{ijk}q^{i}q^{j}q^{k}.
\end{eqnarray}%
The K\"{a}hler moduli space metric is given by 
\begin{equation}
G_{ij}(q)=-\frac{3}{Cqqq}\left( \left( Cq\right) _{ij}-\frac{3\left(
Cqq\right) _{i}\left( Cqq\right) _{j}}{2\left( Cqqq\right) }\right) ,
\label{km1}
\end{equation}%
where we have used the notation

\begin{equation}
Cqqq=C_{ijk}q^{i}q^{j}q^{k},\text{ \ \ \ }\left( Cqq\right)
_{i}=C_{ijk}q^{j}q^{k},\text{ \ }\left( Cq\right) _{ij}=C_{ijk}q^{k}.  \notag
\end{equation}%
Next one has to evaluate the eleven-dimensional Ricci curvature in terms of
the K\"{a}hler moduli taking into consideration that for a $CY_{3},$ we have 
\begin{equation}
g_{\bar{A}\bar{B}}=g_{AB}=R_{AB}=R_{\bar{A}\bar{B}}=R_{A\bar{B}}=0.
\end{equation}%
In addition, we use the Kaluza-Klein ansatz for the three-form 
\begin{equation}
C_{3}=A^{i}\wedge V^{i},
\end{equation}%
then the reduction of the action (\ref{11d}) after a rescaling of the
five-dimensional metric and redefining scalars

\begin{equation}
g_{\hat{\mu}\hat{\nu}}\rightarrow \mathcal{\tilde{V}}^{-\frac{2}{3}}g_{\hat{%
\mu}\hat{\nu}},\text{ \ \ \ \ }h^{i}=\mathcal{\tilde{V}}^{-1/3}q^{i},
\label{sca}
\end{equation}%
gives\footnote{%
We have ignored a kinetic term for the scalar field related to the volume of
the Calabi-Yau and belongs to the hypermultiplet sector.}

\begin{equation}
S_{5}=\int_{M_{5}}\frac{1}{2}R\hat{\ast}1-\frac{1}{2}G_{ij}(h)dh^{i}\wedge 
\hat{\ast}dh^{j}-\frac{1}{4}G_{ij}F_{2}^{i}\wedge \hat{\ast}F_{2}^{j}-\frac{1%
}{12}C_{ijk}A^{i}\wedge F_{2}^{j}\wedge F_{2}^{k},  \label{ra5}
\end{equation}%
where $G_{ij}\left( h\right) $ is given by (\ref{gc}) and $F_{2}^{i}=dA^{i}.$
Note that $G_{ij}\left( h\right) $ is obtained from (\ref{km1}) by simply
replacing the $M^{i}$ with $h^{i}.$

The action of the eleven-dimensional supergravities constructed by Hull \cite%
{Hull} can be written in the form

\begin{equation}
S_{11}=\frac{1}{2}\int_{M_{11}}R\tilde{\ast}1+\frac{\alpha ^{2}}{2}%
F_{4}\wedge \tilde{\ast}F_{4}-\frac{1}{6}C_{3}\wedge F_{4}\wedge \tilde{\ast}%
F_{4},
\end{equation}%
where $\alpha ^{2}=-1$ for the theories with signatures $\left( 1,10\right) $%
, $\left( 5,6\right) $ and $\left( 9,2\right) ,$ and $\alpha ^{2}=1$ for the
mirror theories with signatures $\left( 10,1\right) $, $\left( 6,5\right) $
and $\left( 2,9\right) $. In the reduction of the theories with signatures $%
\left( 1,10\right) $, $\left( 5,6\right) $ and $\left( 2,9\right) ,$ the $%
CY_{3}$ is of signature $\left( 0,6\right) $ and thus $M_{5}$ is of
signature $\left( 1,4\right) $, $\left( 5,0\right) $ and $\left( 2,3\right) .
$ For the reduction of theories with signatures $\left( 10,1\right) $, $%
\left( 6,5\right) $ and $\left( 9,2\right) ,$ the $CY_{3}$ is of signature $%
\left( 6,0\right) $ and thus $M_{5}$ is of signature $\left( 4,1\right) $, $%
\left( 0,5\right) $ and $\left( 3,2\right) .$ All the five-dimensional
supergravity theories obtained have the action

\begin{equation}
S_{5}=\int_{M_{5}}\frac{1}{2}R\hat{\ast}1-\frac{1}{2}G_{ij}(h)dh^{i}\wedge 
\hat{\ast}dh^{j}+\frac{\alpha ^{2}}{4}G_{ij}(h)F_{2}^{i}\wedge \hat{\ast}%
F_{2}^{j}-\frac{1}{12}C_{ijk}A^{i}\wedge F_{2}^{j}\wedge F_{2}^{k}.
\label{ga}
\end{equation}

\section{Four-Dimensional Supergravity}

Starting with the action (\ref{ga}) of the $\mathcal{N}=2$ supergravity
theory in five dimensions with $(t,s)$ signature, we reduce the theory on a
space-like and time-like circle. The Kaluza-Klein reduction ansatz is given
by 
\begin{align}
\mathbf{\hat{e}}^{a}& =e^{-\phi /2}\mathbf{e}^{a},\text{ \ \ \ \ \ \ \ \ }%
\mathbf{\hat{e}}^{0}=e^{\phi }(dt-\sqrt{2}\mathcal{A}^{0}),  \notag \\
A^{i}& =e^{-\phi }x^{i}\mathbf{\hat{e}}^{0}+\sqrt{2}\mathcal{A}^{i},\qquad
h^{i}\ =e^{-\phi }y^{i}.  \label{ra}
\end{align}%
All the fields are taken to be independent of the compact dimension labelled
by index $0$, and the vector $\mathcal{A}^{0}$ has a vanishing component
along the compact dimension. The non-vanishing components of the spin
connection are given by%
\begin{eqnarray}
{\hat{\omega}}_{0,0{\hat{a}}} &=&-\epsilon e^{\frac{\phi }{2}}\partial
_{a}\phi ,  \notag \\
{\hat{\omega}}_{0,{\hat{a}}{\hat{b}}} &=&-{\frac{\epsilon }{\sqrt{2}}}%
e^{2\phi }\mathcal{F}_{ab}^{0},  \notag \\
{\hat{\omega}}_{{\hat{a}},0{\hat{b}}} &=&-{\frac{\epsilon }{\sqrt{2}}}%
e^{2\phi }\mathcal{F}_{ab}^{0},  \notag \\
{\hat{\omega}}_{{\hat{a}},{\hat{b}}{\hat{c}}} &=&e^{\frac{\phi }{2}}\left(
\omega _{a,bc}+{\frac{1}{2}}\delta _{ac}\partial _{b}\phi -{\frac{1}{2}}%
\delta _{ab}\partial _{c}\phi \right) ,  \label{sc}
\end{eqnarray}%
where $\epsilon =1$ corresponds to a reduction on a time-like circle, and $%
\epsilon =-1$ on a space-like circle. Note that all the indices on the right
hand side of (\ref{sc}) are four dimensional, $\omega _{a,bc}$ are the spin
connections of the four-dimensional theory with basis $\mathbf{e}^{a}$ and $%
\mathcal{F}^{0}=d\mathcal{A}^{0}$.

The reduction of the action in (\ref{ga}) results in the following Lagrangian

\begin{eqnarray}
\mathbf{e}^{-1}\mathcal{L}_{4} &=&\frac{1}{2}R-g_{ij}\left( \partial _{{a}%
}x^{i}\partial ^{a}x^{j}+\alpha ^{2}\epsilon \partial _{a}y^{i}\partial
^{a}y^{j}\right)  \notag \\
&&+\frac{\epsilon }{6}Cyyy(g_{ij}\mathcal{F}^{i}.\mathcal{F}^{j}-2\left(
gx\right) _{i}\mathcal{F}^{0}.\mathcal{F}^{i}+\left( gxx\right) \mathcal{F}%
^{0}.\mathcal{F}^{0}+\frac{1}{4}\mathcal{F}^{0}.\mathcal{F}^{0})  \notag \\
&&+\frac{\epsilon }{8}\epsilon ^{abcd}\left( \left( Cx\right) _{ij}\mathcal{F%
}_{ab}^{i}\mathcal{F}_{cd}^{j}-\left( Cxx\right) _{i}\mathcal{F}_{ab}^{0}%
\mathcal{F}_{cd}^{i}+\frac{1}{3}\left( Cxxx\right) \mathcal{F}_{ab}^{0}%
\mathcal{F}_{cd}^{0}\right) .  \label{ract}
\end{eqnarray}%
where 
\begin{eqnarray}
g_{ij} &=&\frac{1}{2}\alpha ^{2}\epsilon e^{-2\phi }G_{ij},\   \notag \\
e^{3\phi } &=&\frac{1}{6}Cyyy.
\end{eqnarray}%
Using (\ref{gc}) and (\ref{ra}) we get 
\begin{equation}
g_{ij}=-\frac{3}{2}\epsilon \alpha ^{2}\left( \frac{\left( Cy\right) _{ij}}{%
\left( Cyyy\right) }-\frac{3\left( Cyy\right) _{i}\left( Cyy\right) _{j}}{%
2\left( Cyyy\right) ^{2}}\right) .  \label{rmetric}
\end{equation}%
In what follows we shall demonstrate that the action (\ref{ract}) describes
a four-dimensional $\mathcal{N}=2$ supergravity theory with various
signatures coupled to vector multiplets with the Lagrangian 
\begin{equation}
\mathbf{e}^{-1}\mathcal{L}=\frac{1}{2}R-g_{ij}\partial _{\mu }z^{i}\partial
^{\mu }\bar{z}^{j}-\frac{\alpha ^{2}}{4}\left( \mathrm{Im}\mathcal{N}_{IJ}%
\mathcal{F}^{I}\cdot \mathcal{F}^{J}+\mathrm{Re}\mathcal{N}_{IJ}\mathcal{F}%
^{I}\cdot \mathcal{\tilde{F}}^{J}\right) ,  \label{fouraction}
\end{equation}%
with the prepotential

\begin{equation}
F=\frac{1}{6}C_{ijk}\frac{X^{i}X^{j}X^{k}}{X^{0}}\ .  \label{pre}
\end{equation}%
The $n$ complex scalar fields $z^{i}$ of $\mathcal{N}=2$ vector multiplets
are coordinates of a projective special (para)-K\"{a}hler manifold. \ In the
symplectic formulation of the theory \cite{vanparis}, one introduces the
symplectic vectors

\begin{equation}
V=\left( 
\begin{array}{c}
X^{I} \\ 
F_{I}%
\end{array}%
\right) ,
\end{equation}%
satisfying the symplectic constraint

\begin{equation}
i_{\epsilon }\left( \bar{X}^{I}F_{I}-X^{I}\bar{F}_{I}\right) =i_{\epsilon
}\left( F_{IJ}-\bar{F}_{IJ}\right) X^{I}\bar{X}^{J}=-N_{IJ}X^{I}\bar{X}%
^{J}=1.  \label{symp}
\end{equation}%
Here $X^{I}=\mathrm{Re}X^{I}+i_{\epsilon }\mathrm{Im}X^{I}$, $i_{\epsilon }$
satisfies $\bar{\imath}_{\epsilon }=-i_{\epsilon }$ and $i_{\epsilon
}^{2}=\tau $, where $\tau =1$ for the case when the scalar fields geometry
is given by a projective special para-K\"{a}hler manifold and $\tau =-1$
when it is given by a projective special K\"{a}hler manifold$,F_{I}=\frac{%
\partial F}{\partial X^{I}}$ and $F_{IJ}=\frac{\partial ^{2}F}{\partial
X^{I}\partial X^{J}}.$ The constraint (\ref{symp}) can be solved by setting 
\begin{equation}
X^{I}=e^{K(z,\bar{z})/2}X^{I}(z)
\end{equation}%
where $K(z,\bar{z})$ is the K\"{a}hler potential. Then we have 
\begin{equation}
e^{-K(z,\bar{z})}=-N_{IJ}X^{I}(z)\bar{X}^{J}(\bar{z})\ .
\end{equation}%
The metric of the special (para)-K\"{a}hler manifold is given by

\begin{equation}
g_{ij}=\frac{\partial ^{2}K(z,\bar{z})}{\partial z^{i}\,\partial {\bar{z}}%
^{j}},  \label{ivp}
\end{equation}%
and locally its $U(1)$ connection $A$ is given 
\begin{equation}
A=-\frac{i_{\epsilon }}{2}(\partial _{i}Kdz^{i}-\partial _{\bar{\imath}}Kd%
\bar{z}^{i}).
\end{equation}%
A convenient choice of inhomogeneous coordinates $z^{i}$ are the special%
\textit{\ }coordinates defined by 
\begin{equation}
X^{0}(z)=1,\text{ \ \ \ }X^{i}(z)=z^{i}\ .
\end{equation}%
Defining 
\begin{equation}
z^{i}=x^{i}-i_{\epsilon }y^{i},  \label{scale}
\end{equation}%
then for theories with cubic prepotentials given in ({\ref{pre}}), we obtain
for the scalars kinetic term 
\begin{equation}
g_{ij}\partial _{\mu }z^{i}\partial ^{\mu }\bar{z}^{j}=\frac{3}{2Cyyy}\tau
\left( \left( Cy\right) _{ij}-\frac{3}{2}\frac{\left( Cyy\right) _{i}\left(
Cyy\right) _{j}}{Cyyy}\right) \left( \partial _{\mu }x^{i}\partial ^{\mu
}x^{i}-\tau \partial _{\mu }y^{i}\partial ^{\mu }y^{i}\right) .
\end{equation}%
The gauge field coupling matrix is given by 
\begin{equation}
\mathcal{\bar{N}}_{IJ}=F_{IJ}(X)+i_{\epsilon }\tau \frac{(N\bar{X})_{I}(N%
\bar{X})_{J}}{\bar{X}N\bar{X}}\;.
\end{equation}%
which for theories with cubic prepotential gives 
\begin{eqnarray}
\mathcal{N}_{00} &=&\frac{1}{3}Cxxx+\tau i_{\epsilon }Cyyy\,\left( \frac{2}{3%
}gxx+\frac{1}{6}\right) ,  \notag \\
\mathcal{N}_{0i} &=&-\frac{1}{2}\left( Cxx\right) _{i}-\frac{2}{3}\tau
i_{\epsilon }\,Cyyy\left( gx\right) _{i},\;  \notag \\
\mathcal{N}_{ij} &=&\left( Cx\right) _{ij}+\frac{2}{3}\tau i_{\epsilon
}g_{ij}\,Cyyy\,.\;  \label{gcm}
\end{eqnarray}%
Using the above information we obtain

\begin{eqnarray}
&&\left( \mathrm{Im}\mathcal{N}_{IJ}\mathcal{F}^{I}\cdot \mathcal{F}^{J}+%
\mathrm{Re}\mathcal{N}_{IJ}\mathcal{F}^{I}\cdot \mathcal{\tilde{F}}%
^{J}\right)   \notag \\
&=&\tau \,Cyyy\left[ -\frac{4}{3}\,\left( gx\right) _{i}\mathcal{F}^{0}\cdot 
\mathcal{F}^{i}\;+\frac{2}{3}g_{ij}\,\,\mathcal{F}^{i}\cdot \mathcal{F}%
^{j}+\left( \frac{2}{3}gxx+\frac{1}{6}\right) \mathcal{F}^{0}\cdot \mathcal{F%
}^{0}\right]   \notag \\
&&+\frac{1}{3}\left( Cxxx\mathcal{F}^{0}\cdot \mathcal{\tilde{F}}%
^{0}-3\left( Cxx\right) _{i}\mathcal{F}^{0}\cdot \mathcal{\tilde{F}}%
^{i}+3\left( Cx\right) _{ij}\mathcal{F}^{i}\cdot \mathcal{\tilde{F}}%
^{j}\right) .
\end{eqnarray}%
After making the identification $\tau =-\alpha ^{2}\epsilon ,$ $\ $and
defining $\mathcal{\tilde{F}}^{iab}=\frac{\tau }{2}\epsilon ^{abcd}\mathcal{F%
}_{cd}^{i},$ (\ref{fouraction}) is equivalent to (\ref{ract}).

Starting in five dimensions with the signatures $(1,4)$, $(3,2)$ and $(5,0)$
and $\alpha ^{2}=-1,$ the reduction on a time-like circle results in
four-dimensional $\mathcal{N}=2$ supergravity theories with signatures $%
(0,4),$ $(2,2)$ and $(4,0)$. The Euclidean supergravity theory (signature $%
(0,4))$ is the one first obtained in \cite{mohaupt3}. The theory of $%
\mathcal{N}=2$ supergravity with $(2,2)$ signature is new and shares some of
the features of the Euclidean theory in the fact that the scalars are
described by a projective special para-K\"{a}hler geometry. The reduction of
the theories with signatures $(1,4)$ and $(3,2)$ on a space-like circle
produces $\mathcal{N}=2$ supergravity theories with signature $(1,3)$ and $%
(3,1)$. These are the well known original theories of $\mathcal{N}=2$
supergravity \cite{vanparis} with projective special K\"{a}hler geometry.
Similarly one obtains $\mathcal{N}=2$ supergravity theories with signatures $%
(0,4),$ $(2,2)$ and $(4,0)$ via the reduction of the theories with $(4,1)$, $%
(2,3)$ and $(0,5)$ signatures on a space-like circle. We also obtain new $%
\mathcal{N}=2$ supergravity theories with signatures $(3,1)$ and $(1,3)$ as
reductions of the five-dimensional theories with signatures $(4,1)$ and $%
(2,3)$ on a time-like circle. These theories have the non-canonical sign of
the gauge fields kinetic terms and have a projective special K\"{a}hler
scalar manifold.

The Killing spinors equations of the five-dimensional supergravity theories
with signatures $(3,2),(1,4)$ and $\left( 5,0\right) $ are given by (\ref%
{ks51}). The reduction of these equations, using the results of \cite{ks},
gives 
\begin{equation}
\left[ {D}_{a}{-}\frac{i}{2}\epsilon \Gamma _{0}A_{a}+\frac{i}{4}%
e^{K/2}\Gamma .\mathcal{F}^{I}\left( \mathrm{Im}X^{J}{(z)}+i\epsilon \Gamma
_{0}\mathrm{Re}X^{J}{(z)}\right) (\func{Im}\mathcal{N})_{IJ}\Gamma _{a}%
\right] \varepsilon =0,
\end{equation}%
and 
\begin{eqnarray}
&&\frac{i}{2}e^{K/2}(\func{Im}\,\mathcal{N})_{IJ}\Gamma .\mathcal{F}^{I}%
\left[ \text{Im}(g^{ij}\mathcal{\bar{D}}_{j}{\bar{X}}^{I}{(z)})+i\epsilon
\Gamma _{0}\text{Re}(g^{ij}\mathcal{\bar{D}}_{j}{\bar{X}}^{I}{(z)})\right]
\varepsilon   \notag \\
&&+\Gamma ^{a}\partial _{a}\left( \func{Re}z^{i}-i\func{Im}z^{i}\Gamma
_{0}\right) \varepsilon   \notag \\
&=&0,
\end{eqnarray}%
where 
\begin{eqnarray}
{D}_{a} &=&\partial _{a}+\frac{1}{4}\omega _{a,bc}\Gamma ^{bc},  \notag \\
A_{a} &=&-\frac{i_{\epsilon }}{2}(\partial _{i}K\partial _{a}z^{i}-\partial
_{\bar{\imath}}K\partial _{a}\bar{z}^{i}),  \notag \\
\mathcal{\bar{D}}_{j}{\bar{X}^{I}(z)} &=&\mathcal{\partial }_{\bar{j}}{\bar{X%
}}^{I}{(z)}+\mathcal{\partial }_{\bar{j}}{K\bar{X}}^{I}{(z).}
\end{eqnarray}%
We also have ${\hat{\varepsilon}=e}^{-\phi /4}\varepsilon ,$ $\left( \Gamma
_{0}\right) ^{2}=-\epsilon $ and $\Gamma ^{0}=-\epsilon \Gamma _{0}$. For $%
\epsilon =1,$ we obtain the Killing spinors for the four-dimensional $%
\mathcal{N}=2$ supergravity theories with $(2,2),(0,4)$ and $\left(
4,0\right) $ while for $\epsilon =-1$ we obtain the Killing spinors for the $%
\mathcal{N}=2$ supergravity theories those of signatures $(3,1)$ and $(1,3).$

The Killing spinors equations of the five-dimensional supergravity theories
with signature $(2,3),(4,1)$ and $\left( 0,5\right) $ are given by

\begin{align}
{{{\hat{D}}}}_{\hat{m}}{\hat{\varepsilon}}-{\frac{1}{8}}h_{i}\left( \Gamma _{%
\hat{m}}{}^{{\hat{n}}_{1}{\hat{n}}_{2}}-4\delta _{{\hat{m}}}^{{\hat{n}}%
_{1}}\Gamma ^{{\hat{n}}_{2}}\right) F_{{\hat{n}}_{1}{\hat{n}}_{2}}^{i}{\hat{%
\varepsilon}}& =0,  \notag \\
\left( F^{i}-h^{i}h_{j}F^{j}\right) _{{\hat{n}}_{1}{\hat{n}}_{2}}\Gamma ^{{%
\hat{n}}_{1}{\hat{n}}_{2}}{\hat{\varepsilon}}+2{\partial }_{\hat{m}%
}h^{i}\Gamma ^{\hat{m}}{\hat{\varepsilon}}& =0.
\end{align}%
Those can be shown to reduce to%
\begin{equation}
\left[ {D}_{a}{\varepsilon +}\frac{1}{2}\epsilon \Gamma _{0}A_{a}{%
\varepsilon }-\frac{1}{4}e^{K/2}\Gamma .\mathcal{F}^{I}\left( \mathrm{Im}%
X^{J}{(z)}-\epsilon \Gamma _{0}\mathrm{Re}X^{J}{(z)}\right) (\func{Im}%
\mathcal{N})_{IJ}\Gamma _{a}\right] \varepsilon =0,
\end{equation}%
and

\begin{eqnarray}
&&-\frac{1}{2}e^{K/2}(\func{Im}\,\mathcal{N})_{IJ}\Gamma .\mathcal{F}^{I}%
\left[ \text{Im}(g^{ij}\mathcal{\bar{D}}_{j}{\bar{X}}^{I}{(z)})-\epsilon
\Gamma _{0}\text{Re}(g^{ij}\mathcal{\bar{D}}_{j}{\bar{X}}^{I}{(z)})\right]
\varepsilon  \notag \\
&&+\Gamma ^{a}\partial _{a}\left( \func{Re}z^{i}-\func{Im}z^{i}\Gamma
_{0}\right) \varepsilon  \notag \\
&=&0.
\end{eqnarray}

For $\epsilon =-1$, we obtain the Killing spinors for four-dimensional $%
\mathcal{N}=2$ superactivities with signatures $(2,2),(4,0)$ and $\left(
0,4\right) $. The Killing spinors for theories with signatures $(1,3),(3,1)$
correspond to $\epsilon =1$.

\section{Summary}

In this work we have constructed $\mathcal{N}=2$ four and five-dimensional
supergravity theories in various space-time signatures. The five-dimensional
theories were constructed by employing the integrability conditions of the
Killing spinor equations as well as by via the reduction of the
eleven-dimensional supergravities constructed by Hull \cite{Hull} on a $%
CY_{3}$. Among the five-dimensional theories constructed, we obtained the
Euclidean five-dimensional supergravity recently constructed in \cite{e5}
and its mirror theory. The four-dimensional supergravity theories were then
obtained as reductions of the five-dimensional theories on a time-like and
space-like circles. One of the new four-dimensional supergravity theories
obtained are the Lorentzian theories with signature $(1,3)$ with projective
special K\"{a}hler geometry and with the wrong sign of the gauge coupling
terms. Solutions of these $(1,3)$ theories with space-like Killing vectors
were considered in \cite{phantom}. There, these theories were labelled as
fake theories. In the present work, however, they were shown to be genuine
theories with higher dimensional origins. Also, in four dimensions a new
theory with signature $\left( 2,2\right) $ is obtained where the scalar
manifold is described by a projective special para-K\"{a}hler manifold. A
future direction is finding solutions to all these theories. The Killing
spinor equations constructed should provide a starting point for a
systematic analysis of their supersymmetric solutions. Also of interest is
the reduction of the four-dimensional theories down to three dimensions and
the investigations of the resulting $c$-maps along the lines of \cite%
{mohaupt4}. We hope to address these questions in forthcoming publications.

\bigskip

\textbf{Acknowledgements} : \  The author would like to thank J. 
Figueroa-O'Farrill for useful discussions. The author also thanks the School
of Mathematics at the University of Edinburgh for hospitality when this work
was completed. This work is supported in part by the National Science
Foundation under grant number PHY-1620505.

\end{document}